# Use of a clinical PET/CT scanner for whole body biodistribution of intranasal nanoparticles.

Michael C Veronesi[1], Marta Zamora[1], Mohammed Bhuiyan[1],
Bill Obrien-Penney[1], Chin Tu Chen[1], and Michael W Vannier[1].

[1]Department of Radiology, University of Chicago, Chicago, IL

**Abstract:**

Whole body biodistribution of 100 nanometer sized polymer micellar nanoparticles (NPs) was determined following intranasal administration using PET/CT imaging on a clinical scanner. Nanoparticles labeled with Zirconium 89 were administered intranasally or intravenously to Sprague Dawley rats followed by serial ex vivo PET/CT imaging in a clinical scanner. At 30 minutes and 1 hour following intranasal delivery, the animals were sacrificed and placed in the PET/CT scanner. Images were acquired and transferred to a workstation for post-processing. 3D regions of interest were constructed. A different set of animals were used for ex vivo verification. These animals were also administered 100 nm polymer micellar NPs and sacrificed at 30 min and 1 hr following intranasal or intravenous delivery via intra-cardiac perfusion. Various organs, including brain, lungs, heart, liver, spleen, stomach and intestines, were procured following exsanguination. A gamma counter determined activity in each organ for comparison with corresponding PET/CT region of interest activity measurements. The ex vivo measurements of activity were consistent with the image-based value determinations. Use of a clinical PET/CT scanner is a feasible means to determine the temporal and spatial distribution of radiolabeled agents after intranasal and intravenous delivery. This work may allow for quantitative in vivo testing of new radionanotheranostic agents.

# 1. Introduction:

Central nervous system (CNS) disorders are often difficult to treat since most drugs cannot cross the blood-brain barrier (BBB). Continued development of alternate pathways for delivery and fate of drugs that can bypass the BBB is needed. Intranasal drug delivery (INDD) is a promising alternative to other forms of drug delivery in its potential for CNS treatment since it is relatively non-invasive, avoids first pass metabolism and side effects can be minimized because smaller drug concentrations are used. Because of its non-invasive nature, there is reduced risk of infection and disease transmission. Nasal spray formulations are easy to administer and can be performed at home by the patient.

So far, a limited number of promising agents have progressed beyond pre-clinical investigation. Even with the drugs that are in clinical trials, the specific mechanism of nose to brain transport is not defined in most cases. Key to overcoming these challenges and furthering the field of INDD is to develop a reliable, informative and non-invasive methodology for better understanding the nose to brain delivery pathway using in vivo advanced imaging both at the clinical and pre-clinical levels. Use of a dedicated small animal positron emission tomography/computed tomography (PET/CT) may not be readily available at some institutions and is costly to install and maintain. An acceptable alternative that has been investigated previously is through the use of a clinical PET/CT (1).

Drug administration and delivery are essential steps in pharmacotherapy (2) (3) (4), where imaging can localize and measure tailored agents such as multifunctional nanoparticles or nanotheranostics to determine their fate (5).

Nanoparticles (NPs) are typically range from 10 to 100 nanometers in diameter and can potentially be targeted at specific molecular sites by surface ligands. Polymer-based NPs can possess optimal characteristics of small size, safety, biodegradability, and customizable surface functionalization that facilitate drug delivery across many biological cell barriers, including the blood brain barrier and the nose-brain barrier (6). Polymer-based NPs have been developed for drug delivery and have received FDA approval for safety (7). When delivered or implanted in the body, PLA, PGA and PLGA undergo hydrolysis and eventually turn into water and carbon dioxide through citric acid cycle (8).

Clinical multimodality imaging of intranasal polymeric nanoparticle (NP) administration via intra-nasal drug delivery (INDD) was used for whole body visualization, localization and measurement in the normal adult Sprague rat. The purpose of this study was to compare intranasal and intravenous whole body biodistribution using PET/CT imaging in rats. A radiolabeled polymeric nanoparticle administered intranasally and intravenously was evaluated at two early time points. This work was done to demonstrate a means for quantitative in vivo pre-clinical testing of new radio-nanotheranostic agents.

# 2. Materials and Methods:

## 2.1. Animals

The protocol for animal experiments was approved by the institutional Animal Care and Use Committee (IACUC). Male, Sprague-Dawley rats (300-500 gm) (Harlan Industries, Indianapolis, IN) were maintained under controlled environmental conditions (23 °C, 12h light/dark cycle) with free access to standard laboratory chow and tap water prior to INDD. A total of 20 animals were utilized for this feasibility study with 6 animals undergoing 2 hr of continuous PET/CT imaging (N=3 for INDD experimental group and N=3 for intravenous control group) and 12 animals undergoing gamma counting following brain isolation and tissue dissection (N=3 for 1 Hr following INDD experimental treatment, N=3 for 1 Hr following IV control treatment, N=5 for 2 hr following INDD experimental treatment and N=3 for 2 hr following IV control treatment).

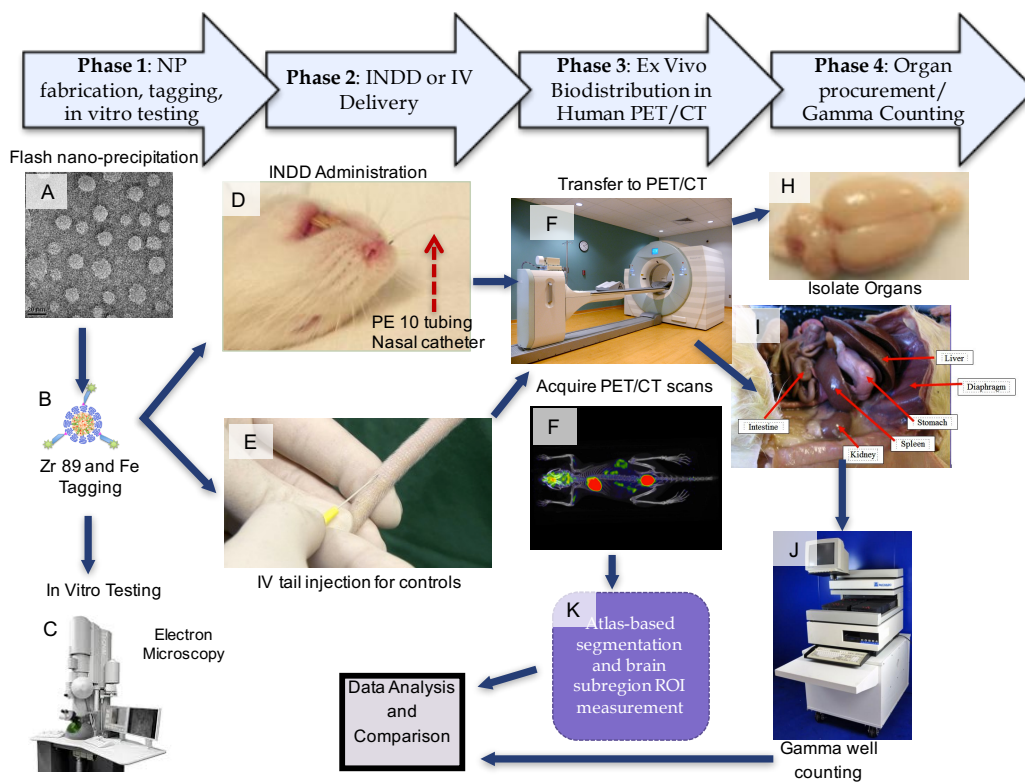

Figure 1. INDD/IV workflow. In Phase 1, the 100 nm PLA-DSPE-PEG NPs were fabricated (A) and tagged with $^{89}$Zr (B) and, underwent morphologic and size characterization in vitro using transmission electron microscopy (C). In phase 2, the NPs were delivered intranasally (D) or intravenously (E) which served as the control comparison. In Phase 3, the rats were euthanized, and transferred to the clinical PET/CT unit (G) where images were acquired and processed (F). Atlas-base segmentation and various organ ROI measurements (brain, lungs, liver, spleen, GI tract, bladder) were obtained. In phase 4, the organs listed above were procured and analyzed in a gamma counter (J). Statistical analysis and comparison was performed between the PET/CT data and gamma data.

## 2.2. Nanoparticles

Polylactic acid (PLA) polymer based nanoparticles (NPs) coated with polyethylene glycoal (PEG) chains containing 20% end amino groups (PLA-PEG NPs) were synthesized (Bio Ma-Tek, Bio Materials Analysis Technology Inc., http://www.bioma-tek.com/, Hsinchu County, Taiwan). The individual compounds used to fabricate the nanoparticles are listed in Table 1. These NPs were characterized in vitro using dynamic light scattering (DLS), zeta potential measurement, and transmission electron microscopy**.** The NP characterizations were performed consistent with ISO 13014 (9). The NPs were shipped dry to our laboratory in Chicago where they were reconstituted into a sterile aqueous suspension prior to experimental use in animals.

Table 1. Nanoparticle Materials

| Material | Company |
|---|---|
| Ammonium thiocyanate | Sigma Aldrich |
| DiR (1,1'-dioctadecyl-3,3,3',3'-tetramethylindotricarbocyanine iodide), PLA [poly(D,L-latide), MW: 75,000-120,000] | Sigma-Aldrich |
| Iron chloride hexahydrate | Sigma-Aldrich |
| Chloroform | J.T. Baker |
| Dichloromethane | J.T. Baker |
| Tetrahydrofuran | J.T. Baker |
| DSPE-PEG2000 {1,2-distearoyl-sn-glycero-3-phosphoethanolamine-N-[methoxy (polyethylene glycol)-2000]} | Avanti Polar Lipids |
| DSPE-PEG2000-NH$_2$ {(1,2-distearoyl-sn-glycero-3-phosphoethanolamine-N-[amino (polyethylene glycol)-2000]) | Avanti Polar Lipids |

## 2.2.1. PLA-PEG nanoparticle preparation

Polylactic acid – polyethylene glycol (PLA-PEG) nanoparticles were prepared as previously reported with modifications (10). Briefly, 5 mg of PLA, 8 mg of DSPE-PEG2000 (1,2-distearoyl-sn-glycero-3-phosphoethanolamine-N-[amino(polyethylene glycol)-2000), 2 mg of DSPE-PEG2000-NH$_2$, and 0.1 mg of DiR were dissolved in 0.5 ml dichloromethane, and then dropped into 3 ml double-distilled water. The mixed solution was emulsified over an ice bath for 1 min using a microtip probe sonicator (XL-2000, Misonix) at 7 W output. The dichloromethane was removed by rotary evaporation to harden nanoparticles. After centrifugation at 13,500 xg for 10 min, pellets were discarded and nanoparticles suspension was washed three times with 10% sucrose solution by 30 kD MWCO ultrafiltration (Vivaspin 6, GE Healthcare). The purified nanoparticles were lyophilized and stored at -20$^0$C. The materials used in nanoparticle handling and their sources are listed in Table 1. The physicochemical characteristics of these nanoparticles were determined and are summarized in Table 2.

Table 2. PLA-PEG NP physiochemical characterization

| Parameter | Result |
|---|---|
| 1.Particle size/size distribution | |
|     Mean diameter (TEM) | 41.1 nm (n=506) |
|     Hydrodynamic diameter (DLS) | 97.1 nm |
|     Polydispersity (DLS) | 0.19 |
| 2.Aggregation/agglomeration state | No aggregation/agglomeration |
| 3.Shape (TEM) | Fluffy sphere (negative stain) |
| 4.Specific surface area | Not applicable |
| 5.Composition (each eppendorf) | 4.67 mg PDA01 + 50 mg sucrose |
|   DiR | 27.1 µg |
|   DSPE-PEG2000 | 3.2 mg |
|   DSPE-PEG2000-$NH_2$ | 64.7 µg (estimated) |
|   PLA | 1.41 mg |
|   Sucrose | 50 mg (estimated) |
| 6.Surface chemistry | PEG-$NH_2$/PEG |
| 7.Surface charge (zeta potential) | -36.0 mV |
| 8.Solubility/dispersibility | ≧9.3 mg PDA / ml (in 10% sucrose) Well dispersed after reconstitution |

## 2.2.2. Transmission electron microscopy (TEM)

Images of nanoparticles were obtained using a transmission electron microscope (TEM, Hitachi model H-7650) using an acceleration potential of 100 kV. Samples were prepared by layering the nanoparticles suspension on a copper grid followed by negative staining for 10 sec with freshly prepared and sterile-filtered 2% (w/v) uranyl acetate solution. The TEM results are depicted in Figure 2 and Table 3.

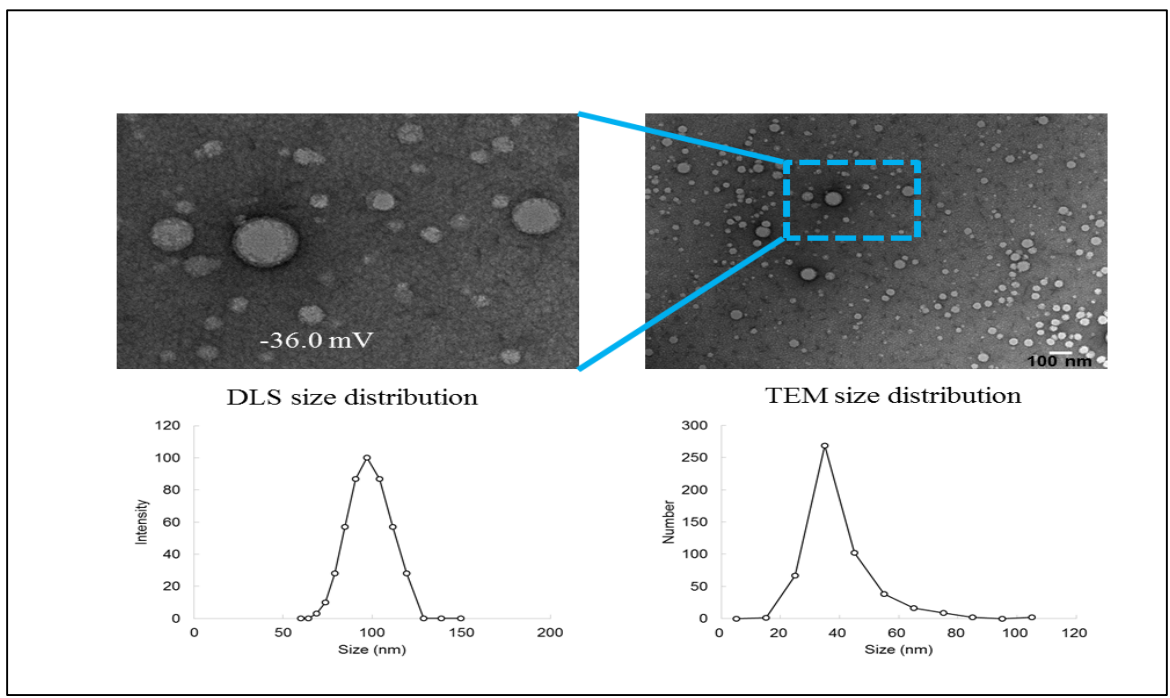

Figure 2: TEM and DLS characterization of PLA-DSPE-PEG NPs. Transmission electron microscopic (TEM) images of PLA-PEG NPs with nominal diameter up to 100 nm corresponding to the scale bar seen in the bottom right corner of the image. The particle size distribution determined by dynamic light scattering (DLS) and TEM are shown. The zeta potential was -36.0 mV.

### 2.2.3. Particle size and zeta potential measurements

Hydrodynamic diameter and zeta potential of PLA-PEG nanoparticles were measured using a particle size analyzer (NanoBrook 90Plus, Brookhaven Instruments Corp., Holtsville NY) and zeta potential analyzer (NanoBrook ZetaPALS, Brookhaven Instruments Corp.) equipped with a 660-nm laser. The measured delay time correlation functions were fitted to a non-negative least squares (NNLS) model to calculate particle size distribution. The DLS results are depicted in Figure 2 and Table 3.

Table 3: Table of Nanoparticle Properties

| System | Surfactant | Indicator | Location of indicator | Mean diameter (nm) | Zeta potential (mv) | Animal | Dose (ug) | Delivery method |
|---|---|---|---|---|---|---|---|---|
| Polymer-micelle (PLA-DSPE-PEG) | None | Zr89 | Covalently linked (Amde linked) | 50-150 | -36 | Sprague-Dawley | 25 | IN or IV |

### 2.2.4. Quantitative determination of DSPE-PEG

DSPE-PEG was measured calorimetrically with ammonium ferrothiocyanate method (11) (15). Samples were dissolved in 1 ml chloroform and mixed with 1 ml of ammonium ferrothiocyanate reagent (30 mg/ml ammonium thiocyanate and 27 mg/ml iron chloride hexahydrate). The mixed solution was shaken for 3 min and centrifuged at 1,000 xg and the red lower layer was collected. The DSPE-PEG derivative was determined at 470 nm absorbance using a spectrophotometer (DU800, Beckman Coulter).

## 2.2.5. PLA quantification

Lyophilized sample was dissolved in stabilized tetrahydrofuran and 20 µl aliquot was analyzed. Chromatographic separation was performed on a gel permeation chromatography system connected to a refractive index detector (Agilent 1100 series) with a PLgel MIXED-D column (300 mm×75 mm, 5 µm, Agilent). PLA was eluted by 100% tetrahydrofuran at a 1 ml/min of flow rate and PLA content was determined by peak area of refractive index signal.

## 2.2.6. Radiolabeling of PLA-PEG NPs with $^{89}$Zr

Zirconium 89 ($^{89}$Zr) was produced with a cyclotron at Washington University at St. Louis and overnight shipped to our institution for nanoparticle tagging. For $^{89}$Zr-labeling, PLA-PEG NPs were first conjugated with a derivative of desferrioxamine (DFO-Bz-NCS) through amide formation. Specifically, 1.55 mg PLA-PEG NPs were stirred with 0.02 mg DFO-Bz-NCS in water for an hour. Purification (molecular weight cut-off (MWCO) 100kD) was done at 8,000 rpm with a centrifugal concentrator (Vivaspin 500 GE HealthCare) and washed with HEPES (4-(2-hydroxyethyl)-1-piperazineethanesulfonic acid) buffer. Before radiolabeling, the PLA-PEG NPs were neutralized with $^{89}$Zr-oxalate by using 1M NaOH in HEPES buffer and a final pH of 7.4 was obtained. Radiolabeled PLA-PEG NPs, $^{89}$Zr (1 mCi) were added to 0.4 mg of PLA-DFO and incubated in pH 7.4 HEPES buffer for 30 minutes. Radiolabeled PLA-PEG NPs were then purified by centrifugation and the labeling efficacy was measured with an instant thin layer chromatography (ITLC) autoradiogram. The radiolabeling activity was 650 µCi per 1 mg of PLA-PEG NPs. Details of zirconium tagging of NPs are listed in Table 4.

Table 4: Zirconium-NP tagging details

| Parameter | Result |
| --- | --- |
| Volume | 1.2 mg NP suspended in 120 ul of 0.9% saline |
| Activity | 967 uCi |
| Percent isolated yield | 60% non-decay corrected yield. |

## 2.7. INDD and IV delivery techniques

For the INDD technique, an adult male rat (300-500g) rat was placed into a 9 x 18 in custom induction chamber and anesthetized with 4% Isoflurane (Figure 3). The animal was weighed and the amount of radioactivity in the nanoparticle solution was measured in a dosimeter. Custom made slots in the side of the induction chamber allowed for manipulation of the animal, which was placed in a supine position. A PE 10 sized nasal catheter was inserted 2 cm deep into the right nostril and attached to a ½ cc Tuberculin syringe. 1.67 µg/1 ul saline of 100 nm PLA-PEG-Zr89 NPs were delivered over 30 s for a total of 25 ug PLA-PEG-Zr89 NPs delivered into the dorsal and posterior aspect of the right nasal cavity. After delivery, a syringe containing NPs was re-measured in a dosimeter to ensure accurate delivery of 100 µCi. For the IV delivery technique, animals were anesthetized using isoflurane via mask ventilation. 3% isoflurane was delivered until complete absence of reflexes. The tail vein was cannulated and 25 ug of NP solution was administered intravenously through a ½ cc Tuberculin syringe slowly manually over several seconds.

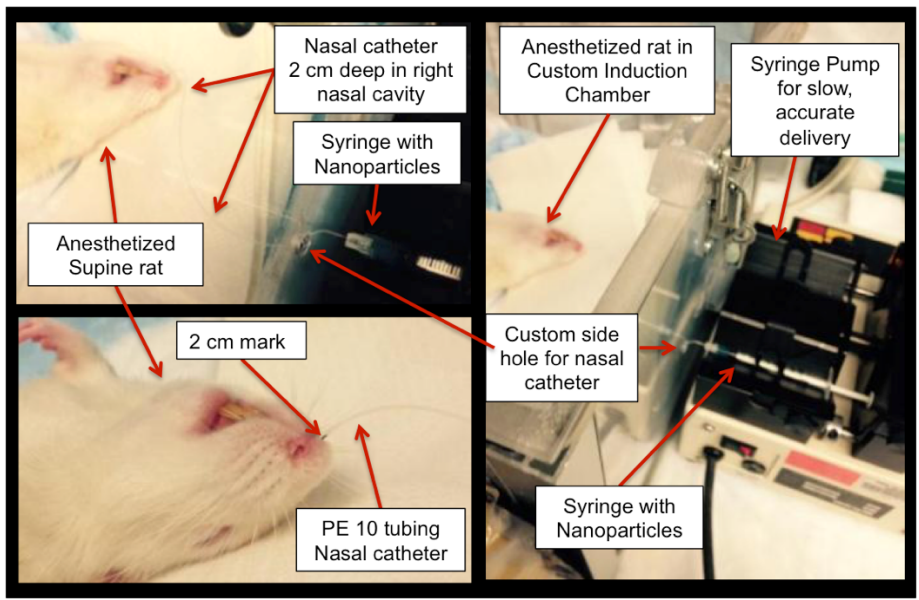
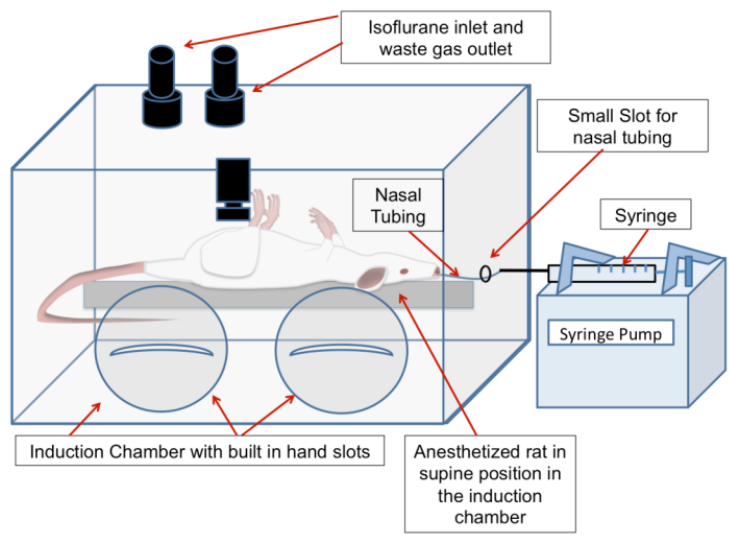

Figure 3: INDD delivery technique. Each rat was placed in an induction chamber, anesthetized and placed in a supine position. Tubing was threaded into the induction chamber through a small custom-made hole drilled on the side and carefully inserted 2 cm deep into the dorsal nasal cavity. The NP solution was delivered via a syringe pump over a 30 second period.

## 2.2.8 PET/CT imaging
The animals were sacrificed using a humane IACUC approved protocol, transported to the human PET/CT scanner and imaged. The PET/CT imaging studies were carried out in a Siemens mCT PET/CT scanner (model: Biograph mCT S (64)-4R Software version VG 51C). Each animal was scanned in two fifteen minute sessions sequentially with the upper half of the animal imaged first followed by the lower half. The FOV for 1 bed was 216 mm, with overlap the 2 beds covered 347 mm. Computer tomography (CT) imaging of animals was performed followed by co-registration of PET images (MIM Version 6.6.5 (MIM Software Inc., Cleveland, OH, http://www.mimsoftware.com/).

**2.2.9 Organ procurement and Gamma counting**

For the organ procurement experiments, 100 nm PLA-DSPE-PEG NPs were labeled with zirconium. 2 mg NP were suspended in 140 ul of 0.9% saline. A 65% non-decay corrected yield was obtained during the labeling. 100 nm PLA-DSPE-PEG NPs were delivered at 30 min and 1 hr following INDD or IV administration. The rats were deeply anesthetized with isoflurane and placed in a dissection tray. The skin of the abdomen, chest and neck shaved and right external jugular vein were exposed. A sub-diaphragmatic approach was followed by open thoracotomy. The Apex of LV was lancinated and feeding needle (+syringe) was placed into ascending aorta. A clamp was placed over feeding needle to prevent reflux of blood from the LV. The right external jugular vein was ligated while saline was pumped through beating heart. 50 ml saline was infused to flush the blood (exsanguination). The organs were dissected, isolated, washed in saline and placed into gamma counting tubes (Brain, liver, lungs, kidneys, spleen, heart, intestines, stomach). Data was collected and entered into an excel sheet and represented as CPM/g weight of each organ.

**3. Results:**

During phase 1 (Figure 1), we produced two sets of NPs (Table 1). The first NP is a 100 nm-sized polymer micellar NP made of Poly-D, L-lactide (PLA), 1,2-distearoyl-sn-glycero-3-phosphoethanolamine (DSPE), and polyethylene gylocol (PEG) to form a 100 nm-sized PLA-DSPE-PEG NP (Bio Ma-Tek, Bio Materials Analysis Technology Inc., http://www.bioma-tek.com/, Hsinchu County, Taiwan). The polymer component of the NP contains the ability to encapsulate a drug agent. The DSPE provides stabilizing properties. The outer PEG contains hydrophilic and stealth properties. For a subset of PEG molecules, an amino group was added to the surface to provide 20% end amino groups for eventual conjugation with tagging agents. The NPs were characterized in vitro using transmission electron microscopy (TEM) and dynamic light scattering (DLS). The mean hydrodynamic diameter of the NPs, which is closest to the true mean diameter in physiologic milieu, was 100 nm with a diameter range of 50-100 nm. The zeta potential was measured to be -36. (Tables 2 and 3 and Figure 2).

The PLA-DSPE-PEG NPs and PBD micellar NPs were prepared for tagging with Zr89 by first conjugating a derivative of desferrioxamine (DFO-Bz-NCS) through amide or carboxyl bonding. Using techniques developed and tested in the RSNA Resident Grant project, Zr89 labeled PLA-PEG NPs were generated and purified by centrifugation. Labeling efficacy for our most recent studies was determined (Approx. 60%) with an instant thin layer chromatography (ITLC) autoradiogram (Table 4).

For phase 2 (Figure 1), 12 animals have been studied to date using either Ex Vivo PET/CT imaging or organ procurement following IV delivery or INDD at various time points. Figure 3 depicts the set up for INDD using a custom-built induction chamber. For intravenous delivery, the tail vein was identified and accessed.

For phase 3 (Figure 1), a total of 8 animals were administered the 100 nm PLA-DSPE-PEG NPs, sacrificed without tissue perfusion or exsanguination and imaged under the clinical mCT PET/CT scanner following NP delivery. PET/CT scanning was consistent in all cases with 15 min/PET acquisitions (2 acquisitions of half body due to crystal array size Using the PET and CT image acquisition and co-registration software (Siemens Biograph mCT Software version VG 51C) images were obtained and fused using MIM6 clinical software (v6.6.5, MIM Software Inc, Cleveland, OH (MIM Software Inc., http://www.mimsoftware.com) Figure 4 shows a representative animal following INDD. Using the MIM6 program, quantitative, three-dimensional regions of interest (3D ROI) were created for select organs which were commonly studied in more recently published nanoparticle biodistribution studies (12) (Figure 5). The whole body (WB) activity as well as brain, nasal cavity, esophagus, stomach, intestines, lungs, liver, spleen, and kidney organ activity were determined in Becquerel/ml after adjusting for animal weight, initial administered dose, and adjusted dose after time decay. The percent activity in the individual organ was calculated based on comparison with whole body activity and these results are depicted for all measured organs in Figure 6 and the brain in Figure 7. Representative images of nasal cavity and brain activity are shown in Figure 8.

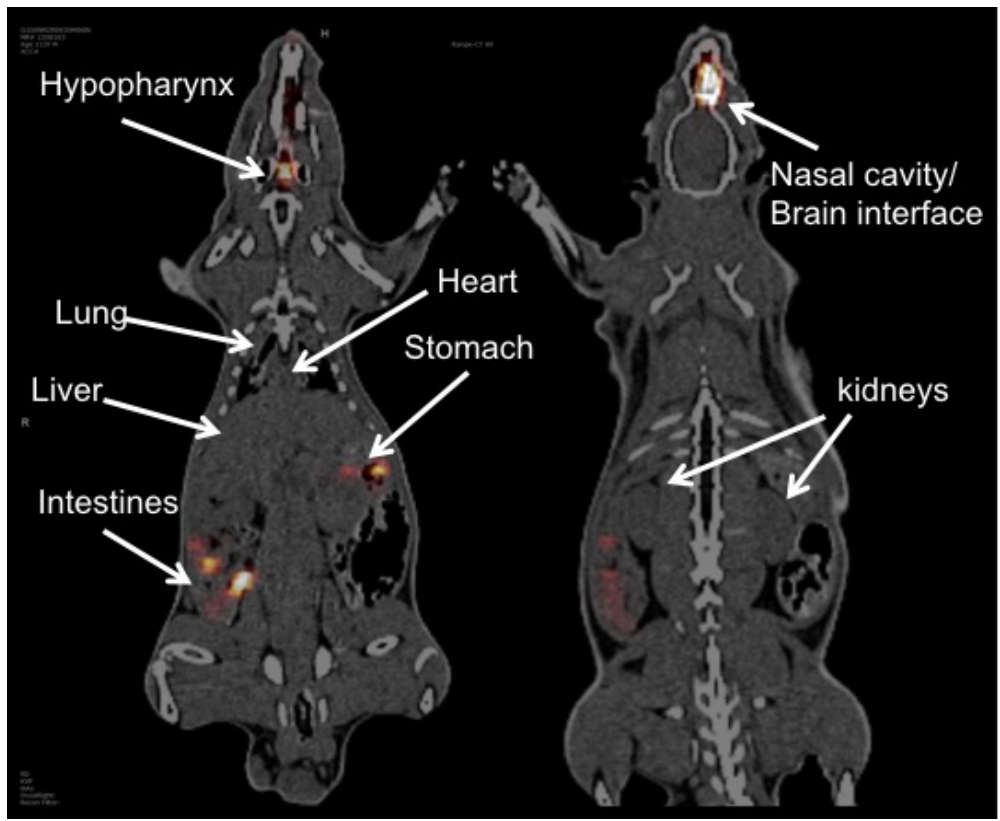

Figure 4: Representative PET/CT images at 1 Hr following INDD of 100 nm PLA-DSPE-PEG NPs tagged with Zr89. Two coronal images showing activity in various organs, most notably the nasal cavity/brain interface, esophagus, stomach and intestines as would be expected post-INDD.

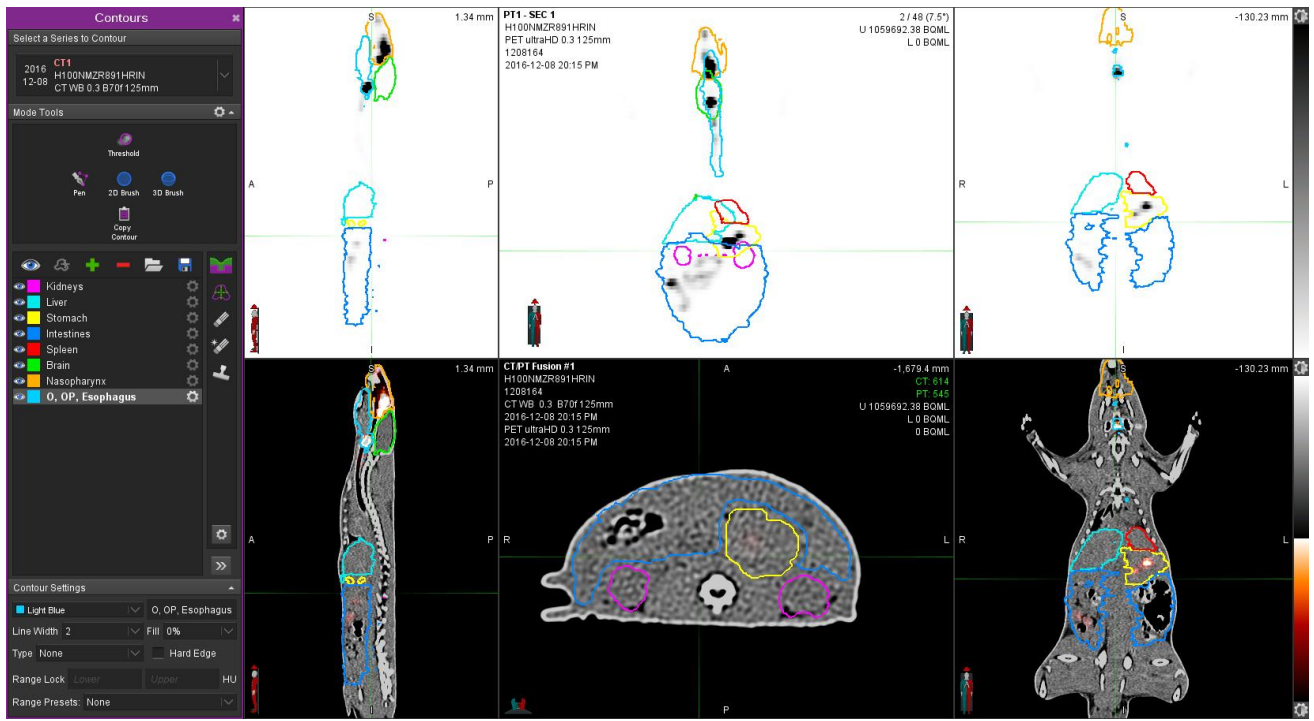

Figure 5: Representative organ region of interest analysis using MIM6 software of a single animal following INDD of 100 nm PLA-DSPE-PEG tagged with Zr89. Organs and their respective colors are shown on the left panel. PET images are shown on top and fusion images of PET with CT are shown on the bottom.

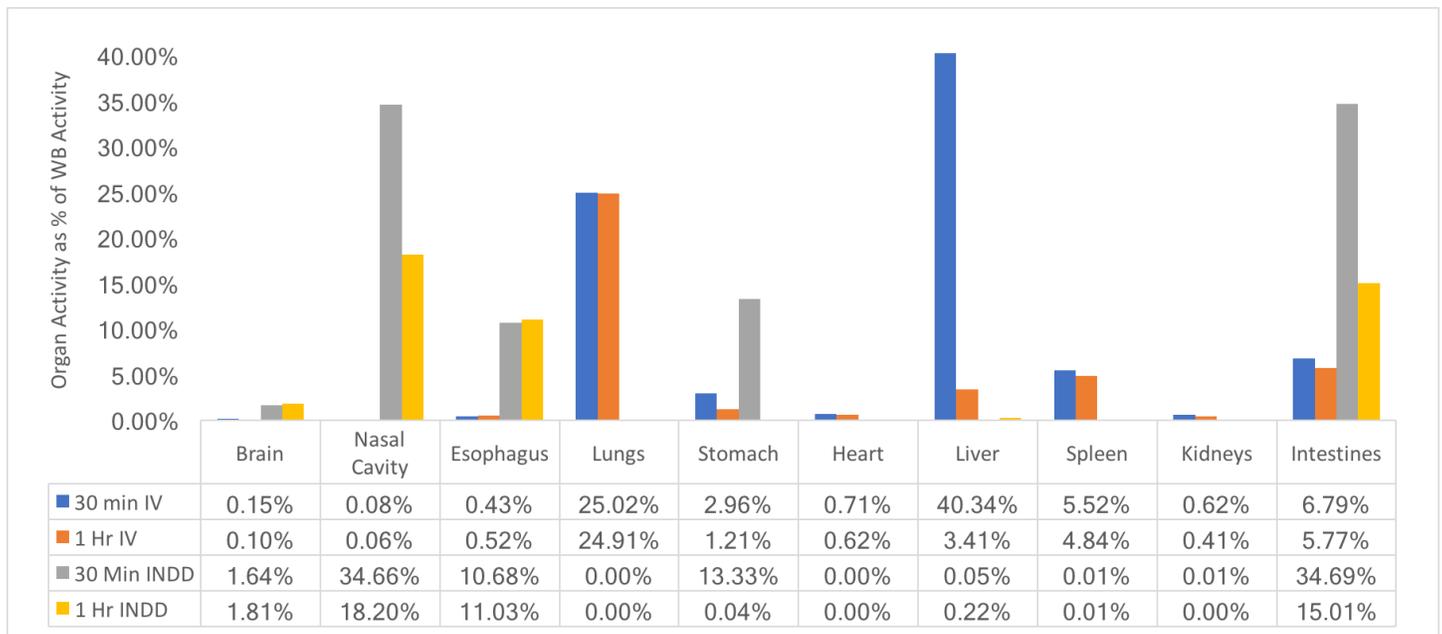

| | Brain | Nasal Cavity | Esophagus | Lungs | Stomach | Heart | Liver | Spleen | Kidneys | Intestines |
|---|---|---|---|---|---|---|---|---|---|---|
| 30 min IV | 0.15% | 0.08% | 0.43% | 25.02% | 2.96% | 0.71% | 40.34% | 5.52% | 0.62% | 6.79% |
| 1 Hr IV | 0.10% | 0.06% | 0.52% | 24.91% | 1.21% | 0.62% | 3.41% | 4.84% | 0.41% | 5.77% |
| 30 Min INDD | 1.64% | 34.66% | 10.68% | 0.00% | 13.33% | 0.00% | 0.05% | 0.01% | 0.01% | 34.69% |
| 1 Hr INDD | 1.81% | 18.20% | 11.03% | 0.00% | 0.04% | 0.00% | 0.22% | 0.01% | 0.00% | 15.01% |

Figure 6: Ex Vivo PET activity in various organs at 30 Min and 1 Hr following intravenous delivery of 100 nm PLA-DPSE-PEG NPs tagged with Zr89. Activity following INDD was greatest as expected in the nasal cavity and along the GI tract. Following IV delivery, activity was greatest in the lung capillary beds, liver, spleen and intestines. Comparatively less signal was seen in the brain, heart and kidneys.

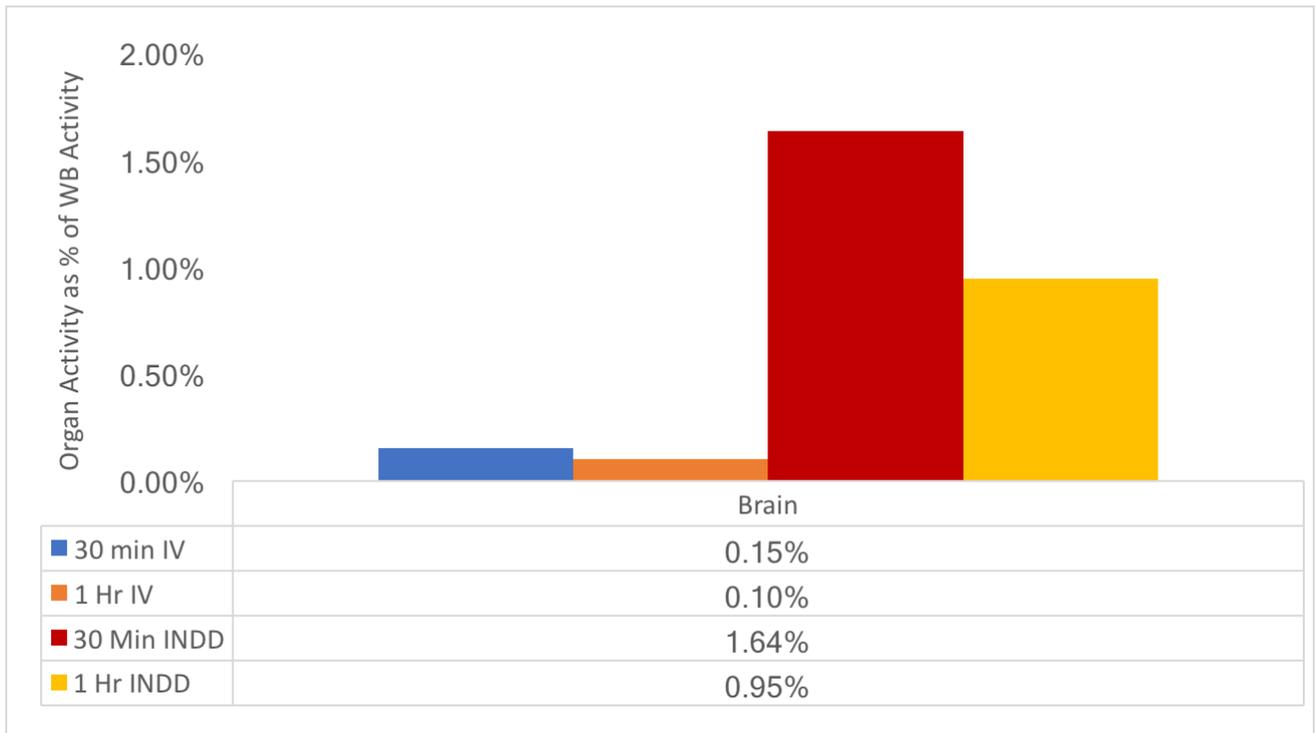

Figure 7: Zr89 activity in the brain at 30 minutes and 1 hour following both IV delivery and INDD of 100 nm PLA-DSPE-PEG NPs. Overall brain activity following IV and INDD was low when depicted as percent of total body activity. However, there was much greater activity in the brain following INDD compared with IV delivery. Brain activity decreased at 1 Hr compared to 30 minutes in both INDD and IV treated animals.

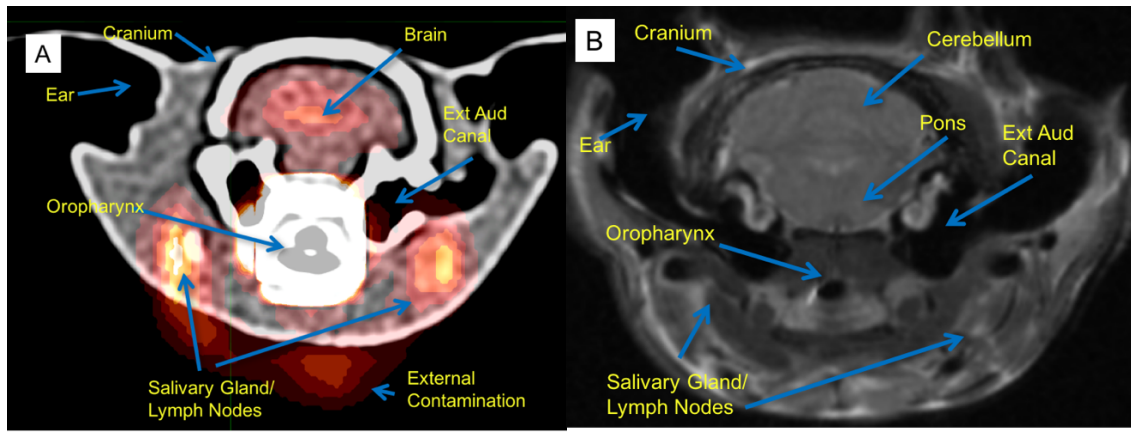

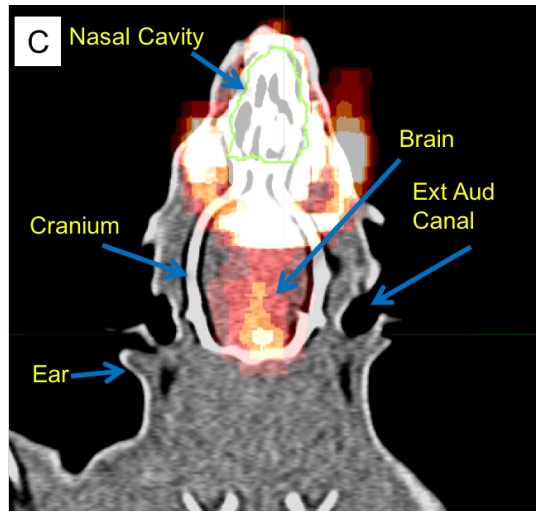

Figure 8: PET activity in the Head/neck and brain in a rat at 30 minutes following administration of 100 nm PLA-DSPE-PEG NPs. A.) Coronal image through the posterior cranium shows intense activity in the oropharynx but also activity in the brain limited to the cranial contour. B.) Coronal MRI of a different rat through the same regions as in image A demonstrates better soft tissue delineation. The brainstem pons and cerebellum as well as the middle cerebellar peduncles are well visualized which corresponds to the region of activity in A. C.) Axial image through the brain, skull and nasal cavity. Note the intense activity in the nasal cavity. Analysis of the anterior brain, including olfactory bulbs is limited given the high signal in the nasal cavity.

During phase 4 (Figure 1), the organs of a total of four animals were procured following both IV delivery and INDD. Animals were sacrificed by intra-cardiac perfusion at either 30 minutes or 1 hour following either IV delivery or INDD. The animals underwent exsanguination via ligation of the internal jugular vein to minimize blood pool contamination of the procured organs. The organs were rinsed in saline, weighed and activity measured in the gamma counting machine. Given the small size of the tube needed for gamma counting, only portions of the stomach and intestines were able to fit into the tubes. Figure 9 demonstrates a similar distribution of activity as seen on PET/CT with the samples represented as counts per minute/gram of tissue (CPM/g tissue).

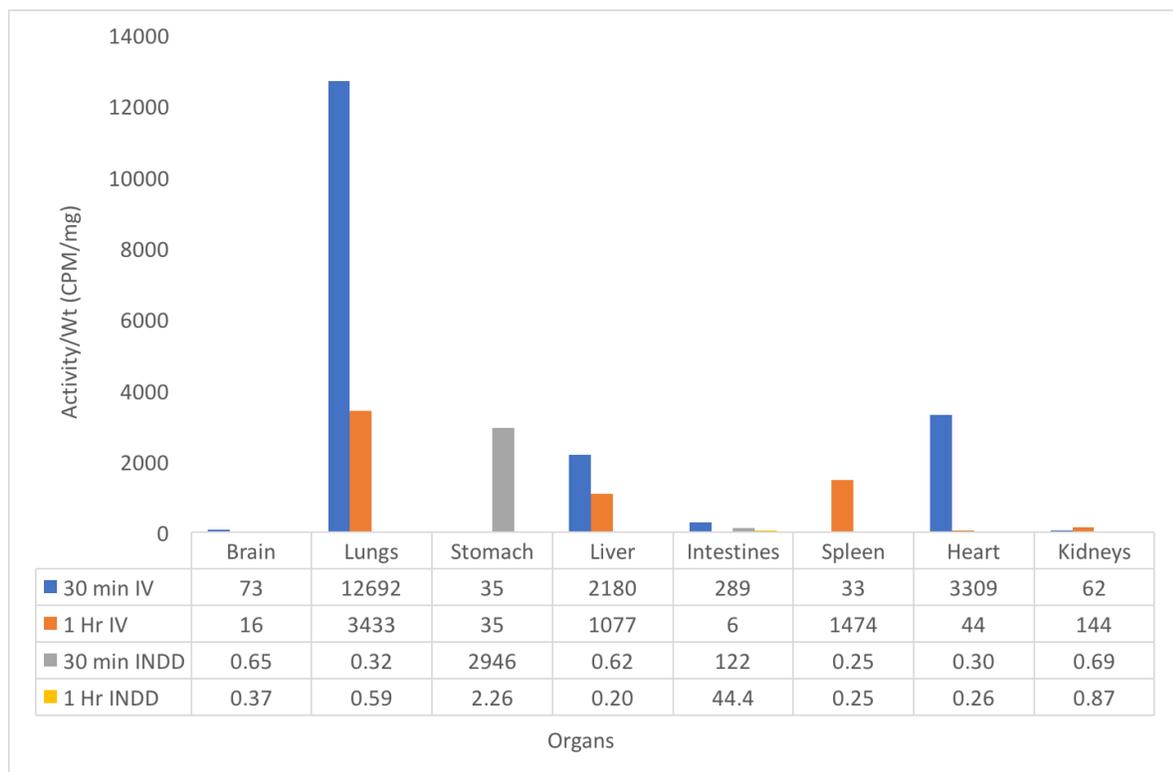

Figure 9: Gamma counter activity in various organs at 30 Min and 1 Hr following intravenous delivery of 100 nm PLA-DPSE-PEG NPs tagged with Zr89. Based on published review by Simon et al., 2014, selected organs included brain, lungs, heart, liver, spleen, kidneys, stomach and intestines. Nasal cavity and esophagus were not measured by gamma counter. In addition, only a portion of the stomach and a portion of the intestines would fit into the gamma counter for measurement.

## 4. Discussion:

The findings reported in this study illustrate that whole body biodistribution can be performed with radiolabeled nanoparticles administered via the intranasal (or intravenous) routes in the ex vivo rat using a clinical PET/CT scanner. Whole body measurements were obtained and compared with initial dose administration and were found to be within 10% of initial administered dose.

By PET/CT, strong activity was visualized in the nasal cavity and pharyngo-esophagus at 30 minutes after INDD, similar to that found by Shingaki et al., 2016 (13). Moderate activity was visualized in the nasal cavity and pharyngo-esophagus at 1 hour after INDD with scant activity appearing in the stomach, liver and intestines. Following organ procurement, intranasally delivered NPs tagged with Zr89, the majority of activity measured within the eight standard organs (not nasal cavity or esophagus) was detected in the stomach at 30 minutes and intestines at 1 hour. This is intra-luminal activity from having been swallowed and is a major limitation of liquid based solutions utilized in INDD. Aerosolized compounds would greatly limit such substantial loss of NPs along the GI tract. Note that activity in the nasal cavity and esophagus were measured during organ procurement since their procurement is technically challenging and readily demonstrated on PET/CT imaging. Contrary to what was expected, only low levels of activity compared to IV delivery were found in the lungs indicating the animal did not aspirate the NPs while under anesthesia. Very low levels were found elsewhere, including the brain, especially after 1 hour.

Based on our data at 30 minute and 1 hour after INDD of 100 nm PLA-DSPE-PEG NPs (Figure 7), from 1-1.5% zirconium activity is visualized the brain, consistent with published literature (14-16). Figure 9 depicts a representative animal imaged 30 min following Ex Vivo imaging with PET/CT. Signal is detected in the posterior half of the brain, although at only the lowest intensity range. Additional much higher signal is seen in the oropharynx and laterally in the anterior neck region corresponding with lymph nodes and/or salivary glands. Strong signal within the nasal cavity precluded assessment of the more anterior brain and olfactory bulbs on this study.

Strong activity was visualized in the liver and lungs following IV delivery. Signal persisted in the lungs 1 hour after IV delivery and declined in the liver substantially. Mild activity was visualized in the spleen and intestines at 30 min following IV delivery. This declined slightly in the spleen and intestines at 1 hour post IV delivery of the NPs. As expected, relatively little activity was seen in the nasal cavity, esophagus, stomach, heart and kidneys at either 30 min or 1 hour after IV delivery. The organs analyzed following intravenous delivery correlate with NP biodistribution Review from Simon et al., 2014 (12). Organs assessed across multiple studies in this review included the brain, liver, kidney, lung, spleen, heart, intestine, and stomach. Following tail vein administration of NPs tagged with Zr89, the majority of gamma counting activity measured within the eight procured organs was detected both the 30 min and 1 hour time points in the lungs, presumably the vascular capillary bed. This correlated with the PET/CT data and suggests possible aggregation or clumping. Very little activity was seen in the brain, stomach, intestines or kidneys following intravenous delivery in the gamma counting studies. The lungs are considered upstream of these organs and could account for low levels in the other organs. Moderate activity was visible in the liver, spleen and blood pool of the heart.

### 4.1 Study Limitations.
Limitations of the study are multiple. This study was performed ex vivo which cannot be performed using list mode format and therefore precludes compartmental analysis. Spatial PET resolution remains a problem when analyzing activity in the anterior brain and skull base regions given their close proximity to the high signal in the posterior nasal cavity and oropharynx. This was also found to be the case with Kamei et al., 2016 as well (17). Brain subregions cannot be delineated with CT so MRI of a reference animal was used to identify brain subregions in Figure 8. Some organs on CT, such as the spleen are not always fully discernable from adjacent organs, such as the stomach and left kidney when determining ROI activity measurement. ROIs definitions are time consuming for individual organs. ROIs from MIM are defined precisely pixel by pixel in 3D with a manual process guided by MRP images of fused PET/CT. Difficulties in ROI analysis arise from Overlap of ROIs since they are not mutually exclusive. The full extent of larger organs, such as the liver, stomach and intestines are amenable to ROI analysis on PET/CT but not using gamma counter where small tubes used for gamma counting allow only limited tissue. This was overcome by using activity/gram of tissue but even with this, gamma counting activity of organs is not directly comparable to the PET/CT activity for these organs.

### Future Work
Many nanoparticles/drugs are being studied using the INDD route but very few have made it to clinical trials. A better understanding of the nose to brain route is essential to maximize delivery of drugs using this route. Since a micro-PET/CT scanner is not available at every research institution, we show the feasibility of using a clinical PET/CT scanner to study INDD. In the future, we plan to study longer time points up to 24 hours after INDD and intravenous delivery to determine the total circulating time of the nanoparticles. Given the overall low rate of compounds entering the brain following intranasal and intravenous delivery, we plan to work with nanoparticle of varying size and zeta potentials. We also will study various agents known to aid in translocation across membranes (Penetration enhancers) and reversibly open tight junctions (chitosan, matrix metalloproteinase inhibitors). We also plan to utilize carbohydrate targeting moieties for better nose to brain delivery (lactoferrin, odoranolectins).

**Conclusion**

We demonstrated that a clinical PET/CT scanner can be utilized to study the whole body biodistribution of radiolabeled nanoparticles administered intranasally to the rat, and can be validated using organ procurement and a gamma counter. Advanced imaging of INDD has the potential to significantly impact the emerging field of radio-nanotheranostics.

**Acknowledgements**

This work was supported by an RSNA Resident Research/Fellow Grants, Hodges Society (University of Chicago) grant, and the National Center for Advancing Translational Sciences of the National Institutes of Health through grant UL1 TR000430. We are especially grateful for technical advice and assistance from Dr. Sean Cheng, Dr. Aaron Tsai, Dr. Lara Leoni, Dr. Leu-wei Lo, Dr. Charles Pelizzari, Ms. Erica Markiewicz, Mr. Jim Vosicky and the late Ms. Charlene Sheridan. The radiotracer for this work was provided by the Division of Radiological Sciences, Washington University School of Medicine, St. Louis MO 63130. For that we thank Dr. Suzzane Lappi, Dr. Dr. Sally Shwartz and Tom Voller for their collaborative efforts. We thank Dr. Chung-Shi Yang, Ph.D. from the Institute of Biomedical Engineering Research and Nanomedicine, National Health Research Institutes, Zhunan, Taiwan 350 for his technical assistance and advice in supplying us with the nanoparticles.